\documentclass[conference]{IEEEtran}

\IEEEoverridecommandlockouts 
\usepackage{booktabs}
\usepackage{textcomp}     

\usepackage{amsmath}
\usepackage{amssymb}
\usepackage{xcolor}
\usepackage{graphicx}
\usepackage[font=small]{caption}
\usepackage[subrefformat=parens]{subcaption}
\usepackage{amsthm}
\usepackage{enumitem}
\usepackage{float}
\usepackage{bm}
\usepackage{enumitem}
\usepackage[noend]{algorithmic}
\usepackage{algorithm}
\usepackage{multirow}
\usepackage{soul}
\usepackage{cite}
\usepackage{mathtools}
\usepackage{xspace}

\theoremstyle{definition}
\newtheorem{example}{Example}

\usepackage{tikz}
\usepackage{pgfplots}
\pgfplotsset{compat=1.18}
\usetikzlibrary{plotmarks}
\usetikzlibrary{matrix,positioning,shapes,arrows,decorations,calc,fit}
\usetikzlibrary{decorations.pathreplacing}
\usetikzlibrary{spy,backgrounds}
\usetikzlibrary{external}
\usetikzlibrary{patterns}
\usetikzlibrary{shapes,arrows}
\usetikzlibrary{shadows.blur}
\usetikzlibrary{shapes.symbols}
\usetikzlibrary{
    pgfplots.groupplots,
    matrix
}

\usetikzlibrary{arrows.meta,positioning,fit,calc}

\usetikzlibrary{calc, arrows.meta}

\IEEEaftertitletext{\vspace{-0.5\baselineskip}}
\begin{document}

\title{Semantic SCL Decoding for Polar Codes}

\author{Andy Huang, Chentao Yue, Branka Vucetic, and Yonghui Li%
\thanks{Code available: https://github.com/huangandy05/Semantic-SCL. The work of Chentao Yue was supported by ARC DECRA under Grant DE250101332.}%
\\
School of Electrical and Computer Engineering, The University of Sydney, Sydney, NSW 2006, Australia\\
E-mail: \{andy.huang1, chentao.yue, branka.vucetic, yonghui.li\}@sydney.edu.au
}

\maketitle

\begin{abstract}
Semantic error control coding (SECC) has emerged as a promising paradigm for improving communication reliability by exploiting the inherent structure and semantic dependencies of transmitted information. This paper proposes a semantic successive cancellation list (SCL) decoder for polar codes carrying natural-language text. A fine-tuned byte-level language model provides next-byte priors, which are converted into conditional bit costs and incorporated directly into the SCL path metric before each pruning step. This allows the decoder to jointly exploit channel reliability and source structure while retaining the conventional polar coding framework. The language model can also be invoked only for bytes whose channel decisions are unreliable, trading decoding performance for latency. Simulations for length-$512$ polar codes show substantial block error rate (BLER) improvements over conventional CRC-aided SCL (CA-SCL) decoding. {At rate $R=0.75$ and list size $L=16$, the proposed decoder achieves more than $1$~dB of gain at a BLER of $10^{-3}$ compared to CA-SCL with the same $L$. Moreover, with $L=8$, it outperforms conventional CA-SCL with $L=256$ in terms of BLER, while requiring about one quarter of the decoding time of the latter in the evaluated software implementation.}
\end{abstract}


\vspace{-0.5em}
\section{Introduction}\label{sec:intro}
\vspace{-0.5em}

Reliable data transmission is a fundamental requirement of modern communication systems, with error-control coding playing a central role in mitigating channel errors. Under the source-channel separation principle \cite{shannon1948}, the channel decoder is designed to treat the information bits as uniform and independent. In practice, however, source data is frequently transmitted with much of its redundancy intact, so the decoder ignores structure still present in the payload. Exploiting this residual structure with learned source models has emerged as a prominent direction within semantic communications \cite{luo2022semcom, gunduz2023beyond}.

Early learning-based approaches to semantic communications largely adopt joint source-channel coding (JSCC), where neural encoder-decoder architectures jointly learn a source representation and its mapping to channel symbols. DeepJSCC maps images directly to channel symbols through a learned end-to-end autoencoder \cite{bourtsoulatze2019deep}, while DeepSC employs a Transformer-based architecture to jointly design semantic and channel coding, optimizing the reconstruction of sentence meaning rather than individual transmitted bits \cite{xie2021deep}. Text-oriented JSCC follows the same end-to-end design~\cite{farsad2018deep}. These approaches are robust to channel noise but replace the source and channel coding pipeline with an end-to-end learned system.

A key challenge of learning-based JSCC is that it does not explicitly separate source and channel coding. Without structured channel coding, it is difficult to provide rigorous performance guarantees for such schemes. To address these issues, semantic error control coding (SECC)~\cite{yue2026secc} retains the existing channel coding scheme at the transmitter while optimizing the decoder to exploit learned semantic information from the source. Language models (LMs) provide such information as a prior over candidate sequences. Hao \emph{et al.}~\cite{hao2025short} use a BART-based model to correct erroneous text segments after decoding independently protected short block codes. More recent work has incorporated LM priors directly into classical decoding algorithms, including Viterbi decoding for convolutional codes and ordered-statistics decoding for short linear block codes \cite{llmviterbi}, \cite{yue2026sem_osd}. These results show that LM priors complement channel reliability information to improve block-error performance without an end-to-end learned system. Nevertheless, none of these LM-aided decoders targets polar codes.

Polar codes~\cite{arikan2009} have been adopted for the control channels of 5G New Radio, and they admit low-complexity decoding. Successive cancellation list (SCL) decoding~\cite{tal2015scl} tracks the $L$ most likely paths, and CRC-aided SCL (CA-SCL)~\cite{niu2012crc} selects the final codeword with an outer CRC.  Both variants prune paths on channel evidence alone. A path discarded at an early information bit cannot be recovered later, and the CRC only selects among the $L$ paths that survive to the end. The pruning step is therefore the stage at which a source prior is most valuable. Wang \emph{et al.}~\cite{wang2016polarlang} exploit this for polar codes carrying Huffman-compressed English text by adding a word-frequency prior from a dictionary to the path metric at every bit and discarding paths that contain words outside the dictionary. A dictionary prior, however, ignores context beyond the current word, and it is confined to the vocabulary it stores.

In this paper, we propose Semantic SCL decoding for polar codes that carry  natural-language text. Its language model, a ByT5 model~\cite{xue2022byt5} fine-tuned on source text, learns a prior over text and supplies it as next-byte probabilities. The decoder converts the probability assigned to each possible next byte into conditional costs for its individual bits. These costs are added, scaled by a weight $\alpha$, to each path's channel path metric before every pruning step, so that paths forming implausible text are pruned. LM evaluations can also be reduced through threshold-based selective guidance. LM-aided re-decoding is invoked only when any channel log-likelihood ratio (LLR) magnitude of a byte falls below a threshold $T$.
We evaluate codes of length $N=512$ with rates $R=0.5$ and $R=0.75$ and a 16-bit CRC. At $R=0.75$, $L=16$, and $E_b/N_0=1.5$~dB, the block error rate (BLER) decreases from $6.31\times10^{-1}$ to $3.68\times10^{-3}$, a factor of approximately $171$. With $L=8$, the proposed decoder outperforms CA-SCL with $L=256$ at about one quarter of its decoding time in our software implementation. Threshold-based guidance reduces the decoding time by about two thirds, at a loss of about $0.2$~dB.
\vspace{-0.3em}
\section{Preliminaries}\label{sec:prelim}
\vspace{-0.3em}


\subsection{System Model}\label{sec:prelim:system}
\vspace{-0.3em}

We consider the transmission of natural-language text. A tokenizer converts the source text into a token sequence $\mathbf{t}=(t_1,t_2,\ldots,t_B)$ of length $B$. In this work, we use the ByT5 language model (see Section~\ref{sec:prelim:byt5-priors}), whose tokens are single bytes, and represent each source character with an 8-bit ASCII encoding, so that tokens, characters and bytes coincide and $t_j\in\{0,\ldots,255\}$. Only these payload bytes are transmitted.

A 16-bit CRC is then appended to the $8B$ payload bits, giving $K=8B+16$ information bits, which are encoded into a length-$N$ polar codeword $\mathbf{c}=(c_0,c_1,\ldots,c_{N-1})$ (see Section~\ref{sec:prelim:polarenc}). The code rate is $R=K/N$ and the effective payload rate is $8B/N$. Note that the proposed framework is not restricted to the use of ASCII encoding or a 16-bit CRC, and other source encoding schemes and CRC configurations can be readily adopted.

\begin{figure}[t]
\centering
\resizebox{\columnwidth}{!}{%
\begin{tikzpicture}[
    >=Latex,
    font=\scriptsize,
    node distance=4mm,
    block/.style={
        draw,
        thick,
        align=center,
        minimum height=5mm,
        inner xsep=2.5pt,
        inner ysep=1.5pt
    },
    arrow/.style={
        ->,
        thick,
        shorten >=1pt,
        shorten <=1pt
    }
]

\coordinate (in) at (0,0);

\node[block, right=2mm of in] (bytecrc)
    {Byte Encoding\\+ CRC};

\node[block, right=of bytecrc] (polar)
    {Polar\\Encoder};

\node[block, right=of polar, minimum width=11mm] (channel)
    {Channel};

\node[block, right=of channel, minimum width=15mm] (decoder)
    {LM-Aided\\CA-SCL};

\node[block, right=of decoder, minimum width=12mm] (text)
    {Byte\\Decoding};

\draw[arrow] ($(in)+(-3mm,0)$) -- (bytecrc);
\draw[arrow] (bytecrc) -- (polar);
\draw[arrow] (polar) -- (channel);
\draw[arrow] (channel) -- (decoder);
\draw[arrow] (decoder) -- (text);

\end{tikzpicture}%
}
\caption{System model for polar-coded text transmission with LM-aided CA-SCL decoding.}
\label{fig:system_model}
\end{figure}
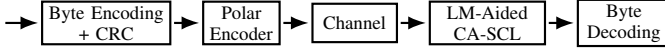

Binary phase shift keying (BPSK) modulation maps the encoded bits ($0\rightarrow +1$ and $1\rightarrow -1$) to the transmitted signal $\mathbf{x}=(x_0,x_1,\ldots,x_{N-1})$\, which is passed through an additive white Gaussian noise (AWGN) channel. The received signal is $\mathbf{y}=\mathbf{x}+\mathbf{n}$, where $\mathbf{n}=(n_0,n_1,\ldots,n_{N-1})$ contains independent and identically distributed (i.i.d.) Gaussian noise with zero mean and variance $\sigma^2=N_0/2$. The signal-to-noise ratio (SNR) is defined as $E_b/N_0$, where $E_b$ is the energy per information bit\, so that $\sigma^2=(2R\,E_b/N_0)^{-1}$ with the CRC bits counted as information bits. The receiver observes $\mathbf{y}$ and estimates $\mathbf{t}$ using the proposed Semantic SCL decoder. Fig.~\ref{fig:system_model} summarizes the system model.


\vspace{-0.3em}
\subsection{Polar Encoding}\label{sec:prelim:polarenc}
\vspace{-0.3em}

Polar codes exploit channel polarization to synthesize bit-channels of differing reliabilities~\cite{arikan2009}. For a code of length $N=2^m$, a set $\mathcal A\subseteq\{0,\ldots,N-1\}$ of $K$ reliable bit-channels is selected to carry information, while the remaining positions $\mathcal A^c$ are frozen to values known to the receiver. The $K$ information bits of Section~\ref{sec:prelim:system}, the payload bytes followed by the CRC, are placed in $\mathcal A$ in increasing index order.

Encoding uses the polar generator matrix $\mathbf G_N = \mathbf B_N \mathbf F^{\otimes m}$, where $\mathbf B_N$ denotes the bit-reversal permutation matrix, $\otimes$ denotes the Kronecker product, and $\mathbf F = \begin{bmatrix}1 & 0 \\ 1 & 1 \end{bmatrix}$ is the basic polarization kernel. Let $\mathbf u=(u_0,\ldots,u_{N-1})$ denote the input vector, where $u_i$ contains an information or CRC bit for $i\in\mathcal A$ and a frozen bit for $i\in\mathcal A^c$. The codeword is then $\mathbf c = \mathbf u \mathbf G_N$, with all operations over $\mathrm{GF}(2)$.

\vspace{-0.3em}
\subsection{CRC-Aided Successive Cancellation List Decoding}\label{sec:prelim:ca-scl}
\vspace{-0.3em}

Successive cancellation (SC) decoding estimates $u_i$, $i=0,\ldots,N-1$, sequentially from the received vector $\mathbf y$ and the preceding decisions~\cite{arikan2009}. A decoding path is a particular sequence of decisions for the previously decoded bits. We denote the $\ell$-th path by $\hat{\mathbf u}_{<i}^{(\ell)}=(\hat u_0^{(\ell)},\ldots,\hat u_{i-1}^{(\ell)})$. For path $\ell$, the channel-only log-likelihood ratio (LLR), evaluated under uniform input bits, is
\begin{equation}
    \lambda_i^{(\ell)}=
    \log
    \frac{\Pr(u_i=0\mid\mathbf y,\hat{\mathbf u}_{<i}^{(\ell)})}
    {\Pr(u_i=1\mid\mathbf y,\hat{\mathbf u}_{<i}^{(\ell)})}.
    \label{eq:sc_llr}
\end{equation}
At $u_i$, SC selects zero for a nonnegative LLR and one otherwise, while frozen bits are fixed to their prescribed values. Since each decision depends on previous decisions, an early error may propagate to later bits~\cite{scflip}.

SCL decoding mitigates this limitation by retaining up to $L$ candidate paths~\cite{tal2015scl}. At each information position, every path branches into two candidates, one for each value $b\in\{0,1\}$, which are ranked using the channel path-metric increment~\cite{llrscl}
\begin{equation}
    \Delta_{\rm ch}^{(\ell)}(i,b)=
    \ln\left(1+\exp\left[-(1-2b)\lambda_i^{(\ell)}\right]\right).
    \label{eq:channel_metric}
\end{equation}
The increment is the negative log-probability of the branch bit given $\mathbf y$ and $\hat{\mathbf u}_{<i}^{(\ell)}$. Each branch inherits its parent's cumulative metric $M_{\rm ch}^{(\ell)}$ and adds~\eqref{eq:channel_metric}, so that $M_{\rm ch}^{(\ell)}$ is the sum of the increments along path $\ell$. Frozen positions are processed without branching using their prescribed bit value. Whenever the number of candidates exceeds $L$, only the $L$ smallest-metric paths are retained. After the final bit, conventional SCL selects the surviving path with the smallest metric. Efficient implementations have complexity $O(LN\log N)$~\cite{tal2015scl}.

CRC-aided SCL (CA-SCL) further improves final path selection by checking the surviving candidates against the appended CRC~\cite{niu2012crc}. Among the CRC-valid paths, the candidate with the smallest metric is selected. If none passes, the smallest-metric path is returned.

\vspace{-0.3em}
\subsection{ByT5 Language-Model Priors}\label{sec:prelim:byt5-priors}
\vspace{-0.3em}

ByT5 is a byte-level encoder-decoder model that operates directly on UTF-8 bytes, avoiding word- or subword-level tokenization~\cite{xue2022byt5}. Its encoder and decoder are internal components of the language model and are unrelated to the polar encoder of Section~\ref{sec:prelim:polarenc} and the channel decoder of Section~\ref{sec:prelim:ca-scl}. The ByT5 encoder extracts contextual representations from the input sequence, while its decoder generates output tokens autoregressively. In this work, we use only the decoder of a ByT5-small model for causal byte-level language modeling, with the ByT5 encoder input fixed to a padding token so that it supplies no context. Given a partial sequence $\mathbf t_{<j}=(t_1,\ldots,t_{j-1})$, the ByT5 decoder outputs a distribution over the next byte $t_j$,
\begin{equation}
    P_\theta(t_j \mid \mathbf t_{<j}),
    \label{eq:byt5_prior}
\end{equation}
where $\theta$ denotes the language-model parameters. The ByT5 decoder produces a logit $z_j(k)$ for every vocabulary index $k$. Since the vocabulary also contains special tokens (e.g., padding, end-of-sequence, and sentinel tokens), we apply softmax normalization over only the $256$ byte values,
\begin{equation}
    P_\theta(t_j = v \mid \mathbf t_{<j})
    = \frac{\exp z_j(\iota(v))}{\sum_{v'=0}^{255}\exp z_j(\iota(v'))},
    \quad v\in\{0,\ldots,255\},
    \label{eq:byte_prior}
\end{equation}
where $\iota(v)$ is the vocabulary index of byte value $v$. The probability of the prefix $\mathbf t_{1:j}$ then follows from the chain rule,
\begin{equation}
    P_\theta(\mathbf t_{1:j}) = \prod_{j'=1}^{j} P_\theta(t_{j'} \mid \mathbf t_{<j'}).
    \label{eq:seq_prob}
\end{equation}
These conditional probabilities quantify the linguistic plausibility of candidate byte sequences and are used as a language-model prior during decoding.

\vspace{-0.3em}
\section{Proposed Semantic SCL Decoder}\label{sec:lmcascl}

\vspace{-0.3em}
\subsection{LM-Aided Path Metric}\label{lm_path_metric}
\vspace{-0.3em}

The proposed decoder follows the CA-SCL procedure in Section~\ref{sec:prelim:ca-scl}, but ranks candidate paths using both channel evidence and a language-model prior. For path $\ell$, the combined metric is
\begin{equation}
 M^{(\ell)}=M_{\rm ch}^{(\ell)}+\alpha M_{\rm LM}^{(\ell)},
 \qquad \alpha\geq0,
 \label{eq:combined}
\end{equation}
where $M_{\rm ch}^{(\ell)}$ accumulates the channel costs in~\eqref{eq:channel_metric}, $M_{\rm LM}^{(\ell)}$ accumulates the language costs derived from the byte prior, defined in Section~\ref{sec:lm:bitcost}, and $\alpha$ controls their relative weight. The language model does not directly observe $\mathbf y$. As in conventional CA-SCL, the decoder retains the $L$ candidates with the smallest metrics whenever the list exceeds $L$. For $\alpha=0$, LM inference is bypassed and conventional CA-SCL decoding is used.

\vspace{-0.3em}
\subsection{Byte Priors and Bit Costs}\label{sec:lm:bitcost}
\vspace{-0.3em}

The language model predicts one byte at a time, while the list decoder prunes after every bit. A byte spans eight of these prunings. If the language cost were applied only at the byte boundary, all eight would rank paths by the channel metric alone. This section builds $M_{\rm LM}^{(\ell)}$ by assigning a language cost to each of the eight bits inside a byte, all obtained from a single call to the language model.

At the start of byte $j$, the previously decoded bytes $\hat{\mathbf t}_{<j}^{(\ell)}=(\hat t_1^{(\ell)},\ldots,\hat t_{j-1}^{(\ell)})$ of each surviving path $\ell$ are supplied to ByT5 to obtain the next-byte distribution $p_{j,\ell}(v)=P_\theta(v\mid\hat{\mathbf t}_{<j}^{(\ell)})$, $v=0,\ldots,255$, where for $j=1$ the prefix is empty and the ByT5 decoder receives only its start token. The paths are evaluated together in a batch. The same distribution is then used throughout the next eight payload-bit decisions.

To use this byte distribution during bitwise pruning, consider a candidate byte whose first $r$ bits are $\mathbf h_{1:r}^{(\ell)}$, in most-significant-bit-first order. Let $\mathcal V(\mathbf h)$ be the set of byte values whose binary representations begin with $\mathbf h$. Their total probability is
\begin{equation}
 Q_{j,r}^{(\ell)}
 =\sum_{v\in\mathcal V(\mathbf h_{1:r}^{(\ell)})}p_{j,\ell}(v),
 \qquad Q_{j,0}^{(\ell)}=1.
 \label{eq:mass}
\end{equation}
For the branch choosing the $r$th bit, $r=1,\ldots,8$, the additional language cost is
\begin{equation}
 \Delta_{\rm LM}^{(\ell)}(j,r)
 =-\log\frac{Q_{j,r}^{(\ell)}}{Q_{j,r-1}^{(\ell)}}.
 \label{eq:lm_increment}
\end{equation}
The ratio is the probability of the new bit conditioned on the preceding bytes and the already chosen bits of the current byte. The language term $M_{\rm LM}^{(\ell)}$ of~\eqref{eq:combined} accumulates these costs along the path.
The eight costs sum to the negative log-probability of the completed byte:
\begin{equation}
 \sum_{r=1}^{8}\Delta_{\rm LM}^{(\ell)}(j,r)
 =-\log P_\theta(\hat t_j^{(\ell)}
                 \mid\hat{\mathbf t}_{<j}^{(\ell)}).
 \label{eq:telescope}
\end{equation}
By~\eqref{eq:telescope} and the chain rule~\eqref{eq:seq_prob}, a complete path with payload $\hat{\mathbf t}^{(\ell)}=(\hat t_1^{(\ell)},\ldots,\hat t_B^{(\ell)})$ has
\begin{equation}
 M_{\rm LM}^{(\ell)}=-\log P_\theta(\hat{\mathbf t}^{(\ell)}).
 \label{eq:lm_metric}
\end{equation}

The ratio in~\eqref{eq:lm_increment} is well defined because the renormalization in~\eqref{eq:byte_prior} leaves every byte value with positive probability, so $Q_{j,r}^{(\ell)}>0$ for every prefix.

Both terms of~\eqref{eq:combined} are negative log-probabilities. By~\eqref{eq:lm_metric}, $M_{\rm LM}^{(\ell)}$ is the negative log-prior of the payload. Since~\eqref{eq:sc_llr} assumes uniform input bits, $M_{\rm ch}^{(\ell)}$ is the negative log-likelihood of the path. With $\alpha=1$, ranking by~\eqref{eq:combined} is therefore the maximum a posteriori rule for a source with prior $P_\theta$, and $\alpha<1$ discounts $P_\theta$ when it is overconfident relative to the true source distribution.

\vspace{-0.3em}
\subsection{Decoding Procedure}\label{sec:lm:procedure}
\vspace{-0.3em}

A byte is termed \emph{guided} when the LM is evaluated before its decoding.

Each branch inherits its parent's metric and receives the update
\begin{equation}
 M^{(\ell)}\gets M^{(\ell)}
 +\Delta_{\rm ch}^{(\ell)}(i,b)+\alpha\delta_i^{(\ell)},
 \label{eq:metric_update}
\end{equation}
where $\delta_i^{(\ell)}$ is given
by~\eqref{eq:lm_increment} for payload bits and is zero for frozen and CRC bits. After decoding, the CRC-valid survivor with the smallest combined metric is selected, or the smallest-metric survivor if none passes. We refer to this decoder as every-byte Semantic SCL.

\vspace{-0.3em}
\subsection{Threshold-Based LM Guidance}\label{sec:lm:threshold_semscl}
\vspace{-0.3em}

To reduce LM invocations, we propose guiding bytes based on channel reliability. Before decoding payload byte $j$, the decoder saves its state and provisionally decodes the eight payload bits without new LM costs, while retaining previously accumulated costs. Let $\widetilde{\mathcal L}_j$ denote the provisional survivors with combined metrics $\widetilde M^{(\ell)}$. The reliability of the best survivor is measured as
\begin{equation}
\begin{split}
 \ell_j^\star &= \arg\min_{\ell\in\widetilde{\mathcal L}_j}
 \widetilde M^{(\ell)},\\
 \rho_j &= \min_{r=1,\ldots,8}
 \left|\widetilde\lambda_{j,r}^{(\ell_j^\star)}\right|,
\end{split}
\label{eq:byte_reliability}
\end{equation}
where $\widetilde\lambda_{j,r}^{(\ell)}$ is the LLR~\eqref{eq:sc_llr} of payload bit $r$ of byte $j$ along path $\ell$ during provisional decoding.

If $\rho_j<T$, where $T\geq0$ is a predefined threshold, the decoder restores the pre-byte state and re-decodes the byte with LM guidance, applying the conditional bit costs before each pruning step. Otherwise, the provisional results are retained without additional LM costs. The resulting set of guided bytes is $\mathcal S_T=\{j:\rho_j<T\}$, giving the accumulated LM cost
\begin{equation}
 M_{\rm LM}^{(\ell)}=-\sum_{j\in\mathcal S_T}
 \log P_\theta\big(\hat t_j^{(\ell)}\mid
 \hat{\mathbf t}_{<j}^{(\ell)}\big).
 \label{eq:selective_lm_metric}
\end{equation}
Restoration discards provisional updates, ensuring that costs are accumulated only once.

Lowering $T$ restricts LM guidance to bytes containing more uncertain bits, reducing LM invocations and latency at the expense of error-rate performance. At $T=0$, the decoder reduces to conventional CA-SCL, and $T=\infty$ yields every-byte Semantic SCL. Algorithm~\ref{alg:decoder} summarizes the proposed decoding procedure.

\begin{algorithm}[h]
\caption{Semantic SCL Decoding}
\label{alg:decoder}
\small
\begin{algorithmic}[1]
\REQUIRE Received vector $\mathbf y$, $N$, $\mathcal A$, payload length $B$,
list size $L$, byte model $P_\theta$, threshold $T\geq0$, weight $\alpha\geq0$
\ENSURE Decoded payload $\hat{\mathbf t}$
\STATE Initialize the CA-SCL paths and channel metrics; set all LM costs to zero
\FOR{$i=0,\ldots,N-1$}
  \IF{$i$ is the first payload bit of byte $j$}
    \STATE Save the list state, provisionally decode byte $j$ without new LM costs, and compute $\rho_j$ by~\eqref{eq:byte_reliability}
    \STATE If $\rho_j\geq T$, keep this result and continue after byte $j$; otherwise, restore the list state
    \STATE Obtain and store $p_{j,\ell}$ for each path using~\eqref{eq:byte_prior}
  \ENDIF
  \STATE Extend paths as in CA-SCL, branching on both bits if $i\in\mathcal A$ and otherwise taking the frozen value
  \STATE Each extension inherits its parent's metric and stored byte distribution
  \STATE Compute the channel increment using~\eqref{eq:channel_metric}
  \STATE Set $\delta_i$ using~\eqref{eq:lm_increment} for a guided bit, otherwise zero
  \STATE Update every candidate metric using~\eqref{eq:metric_update}
  \STATE Retain up to $L$ smallest-metric paths, together with their LM state
\ENDFOR
\STATE Form $\mathcal L_{\rm CRC}$ from the CRC-valid survivors
\STATE If $\mathcal L_{\rm CRC}$ is empty, use all survivors instead
\RETURN payload of the smallest-metric path in $\mathcal L_{\rm CRC}$
\end{algorithmic}
\end{algorithm}

\vspace{-0.3em}
\subsection{Language Model Adaptation}\label{sec:byt5_finetuning}
\vspace{-0.3em}
ByT5 is pretrained to reconstruct masked spans of its encoder input~\cite{xue2022byt5}, whereas the ByT5 decoder in Section~\ref{sec:prelim:byt5-priors} receives no encoder context. We therefore fine-tune the ByT5 decoder of the pretrained ByT5-small model for next-byte prediction on source-domain text. We select 10,000 sentences with lengths between 80 and 120 characters from the training set of the Stanford Natural Language Inference (SNLI) corpus~\cite{bowman2015large} and split them into a training set $\mathcal D_{\rm tr}$ of 9000 sentences and a validation set of 1000 sentences.

Training uses teacher forcing, with the ByT5 encoder input fixed to a padding token as in Section~\ref{sec:prelim:byt5-priors}. All ByT5 decoder parameters are fine-tuned by minimizing the next-byte negative log-likelihood
\begin{equation}
\mathcal J(\theta)
=
-\sum_{\mathbf t\in\mathcal D_{\rm tr}}
\sum_{j=1}^{|\mathbf t|}
\log P_\theta
\left(t_j\mid\mathbf t_{<j}\right),
\label{eq:training}
\end{equation}
where $|\mathbf t|$ is the number of bytes in $\mathbf t$. During training, $P_\theta$ is normalized over the full ByT5 vocabulary rather than over the byte values as in~\eqref{eq:byte_prior}, and the loss also includes the end-of-sequence token that follows $t_{|\mathbf t|}$. Because the loss is evaluated at every byte position, the same model serves any threshold $T$ in Section~\ref{sec:lm:threshold_semscl}. Fine-tuning uses AdamW with learning rate $5\times10^{-5}$, weight decay 0.01, and batch size 32. The learning rate decays linearly after a 5\% warm-up, and gradients are clipped at norm 1. Training runs for up to ten epochs and stops early once the validation loss has not improved for two consecutive epochs.

\begin{figure}[h]
\centering
\captionsetup[sub]{skip=3pt}
\colorlet{lmtarget}{blue!14}
\colorlet{lmtargettext}{blue!75!black}
\colorlet{lmkeep}{gray!7}
\colorlet{lmdrop}{gray!30}
\colorlet{lmdroptext}{gray!60}
\tikzset{
  lmpick/.style={draw=black, line width=0.5pt, rounded corners=1.5pt,
                 inner sep=1.5pt},
  lmrule/.style={line width=0.3pt},
}
\begin{subfigure}{\columnwidth}
\centering
\begin{tikzpicture}[x=0.68cm, y=0.42cm, font=\footnotesize,
  >={Latex[length=3.5pt,width=3pt]}]
\draw[->, gray!70] (0.55,1.0) -- (8.45,1.0)
  node[midway, above=-1pt, text=black]
  {bit $r$ of byte $j$, in decoding order};
\foreach \rr in {1,...,8} \node at (\rr,0) {$\rr$};
\node[anchor=east] at (0.35,0) {$v$};
\node at (9.65,0) {$p_{j,\ell}(v)$};
\draw[line width=0.6pt] (-1.65,-0.6) -- (10.4,-0.6);

\foreach \yy/\vv/\bitlist/\pv/\dc/\tg in {
    1/b/{0,1,1,0,0,0,1,0}/0.25/6/0,
    2/f/{0,1,1,0,0,1,1,0}/0.10/8/0,
    3/g/{0,1,1,0,0,1,1,1}/0.40/9/1,
    4/p/{0,1,1,1,0,0,0,0}/0.05/4/0,
    5/s/{0,1,1,1,0,0,1,1}/0.20/4/0}{
  \ifnum\tg=1
    \def\rowfill{lmtarget}\def\rowtext{lmtargettext}\def\rowfont{\bfseries}
  \else
    \def\rowfill{lmkeep}\def\rowtext{black}\def\rowfont{}
  \fi
  \node[anchor=east, text=\rowtext] at (0.35,-\yy) {\texttt{\rowfont\vv}};
  \node[text=\rowtext] at (9.65,-\yy) {\rowfont\pv};
  \foreach \bit [count=\rr] in \bitlist {
    \ifnum\rr<\dc
      \def\cellfill{\rowfill}\def\celltext{\rowtext}
    \else
      \def\cellfill{lmdrop}\def\celltext{lmdroptext}
    \fi
    \fill[\cellfill, rounded corners=1pt]
      (\rr-0.43,-\yy-0.39) rectangle (\rr+0.43,-\yy+0.39);
    \node[text=\celltext] at (\rr,-\yy) {\rowfont\texttt{\bit}};
  }
}

\draw[lmrule] (-1.65,-5.65) -- (10.4,-5.65);
\node[anchor=east] at (0.35,-6.45) {$Q_{j,r}^{(\ell)}$};
\node[anchor=east] at (0.35,-7.65) {$\Delta_{\rm LM}^{(\ell)}(j,r)$};
\foreach \qv/\dv/\hl [count=\rr] in {
    1.00/0.00/0, 1.00/0.00/0, 1.00/0.00/0, 0.75/0.29/1,
    0.75/0.00/0, 0.50/0.41/1, 0.50/0.00/0, 0.40/0.22/1}{
  \node at (\rr,-6.45) {$\qv$};
  \ifnum\hl=1
    \node[text=lmtargettext] at (\rr,-7.65) {$\boldsymbol{\dv}$};
  \else
    \node[text=lmdroptext] at (\rr,-7.65) {$\dv$};
  \fi
}
\end{tikzpicture}
\caption{Candidate bytes after \texttt{A dog runs on the\textvisiblespace}}
\end{subfigure}

\vspace{0.5em}
\begin{subfigure}{\columnwidth}
\centering
\begin{tikzpicture}[x=0.68cm, y=0.45cm, font=\footnotesize]
\node at (3.5,1.95) {$\Delta_{\rm ch}^{(\ell)}(i,b)$};
\node at (5.75,1.95) {$\Delta_{\rm LM}^{(\ell)}(j,4)$};
\node at (7.9,1.95) {sum};
\draw[lmrule] (2.55,1.45) -- (8.6,1.45);

\coordinate (par) at (0,0);
\node[anchor=east] at (-0.2,0) {$\lambda_i^{(\ell)}=-0.8$};
\node[draw=lmtargettext, fill=lmtarget, text=lmtargettext,
      rounded corners=2pt, inner sep=2pt] (c0) at (1.35,0.8) {$b=0$};
\node[draw=gray!60, fill=lmkeep, rounded corners=2pt, inner sep=2pt]
      (c1) at (1.35,-0.8) {$b=1$};
\draw[lmtargettext, line width=0.7pt] (par) |- (c0.west);
\draw[gray!70, line width=0.5pt] (par) |- (c1.west);
\fill[black] (par) circle (1.7pt);

\node[text=lmtargettext] at (3.5,0.8) {$1.17$};
\node at (4.6,0.8) {$+$};
\node[text=lmtargettext] at (5.75,0.8) {$0.29$};
\node at (6.85,0.8) {$=$};
\node[lmpick, text=lmtargettext] at (7.9,0.8) {$\boldsymbol{1.46}$};
\node[lmpick] at (3.5,-0.8) {$0.37$};
\node at (4.6,-0.8) {$+$};
\node at (5.75,-0.8) {$1.39$};
\node at (6.85,-0.8) {$=$};
\node at (7.9,-0.8) {$1.76$};
\end{tikzpicture}
\caption{Branch increments at the fourth bit}
\end{subfigure}
\caption{Bit costs in Example~\ref{ex:bitcost}. Shaded bits depart from \texttt{g}, and boxes mark the smaller channel cost and the smaller sum.}
\label{fig:lm_example}
\end{figure}

\begin{example}\label{ex:bitcost}
Let the payload contain the sentence \texttt{A dog runs on the grass}, and let byte $j$, the \texttt{g} of \texttt{grass}, be guided. Suppose that path $\ell$ has correctly decoded all bytes before it, and let $p_{j,\ell}$ assign $0.40$, $0.25$, $0.20$, $0.10$, and $0.05$ to \texttt{g}, \texttt{b}, \texttt{s}, \texttt{f}, and \texttt{p}, and negligible probability to the other byte values. Fig.~\ref{fig:lm_example}(a) shows the costs in Eq. ~\eqref{eq:lm_increment} along \texttt{g}. They are nonzero only where candidates are excluded: the fourth bit excludes \texttt{p} and \texttt{s} and costs $0.29$, the sixth excludes \texttt{b} and costs $0.41$, and the eighth excludes \texttt{f} and costs $0.22$. Their sum is $-\log0.40=0.92$, as~\eqref{eq:telescope} requires.

Suppose further that the fourth bit $u_i$ has $\lambda_i^{(\ell)}=-0.8$. For $b=0$ and $b=1$, the channel costs~\eqref{eq:channel_metric} are $1.17$ and $0.37$, and the language costs are $0.29$ and $-\log0.25=1.39$. With $\alpha=1$, the increments in~\eqref{eq:metric_update} are $1.46$ and $1.76$, as in Fig.~\ref{fig:lm_example}(b). The correct branch $b=0$ is thus ranked first, although the channel alone favors $b=1$. Applied only after the eighth bit, the language cost would come too late if the correct branch were pruned here.
\end{example}

\vspace{-0.5em}
\section{Experiments}\label{sec:exp}
\vspace{-0.3em}
\subsection{Setup}\label{sec:exp:setup}
\vspace{-0.3em}
We evaluate the proposed decoder using two polar-code configurations with block length $N=512$ and code rates $R=0.75$ and $R=0.5$. Source sequences are drawn from the premise and hypothesis fields of the SNLI test set~\cite{bowman2015large}, retaining samples of at least 80 bytes and truncating each sequence to its first $B=(NR-16)/8$ bytes to fit the polar payload, namely $B=46$ for $R=0.75$ and $B=30$ for $R=0.5$. The baseline is conventional CA-SCL with the same polar code and CRC. It uses list sizes $L\in\{4,16,256\}$, and the proposed decoder uses $L\in\{4,8,16\}$ with every-byte guidance except in Section~\ref{sec:exp:threshold}.

The pretrained ByT5-small model is fine-tuned as described in Section~\ref{sec:byt5_finetuning} using a single NVIDIA GeForce RTX 5090 GPU. For each code configuration and $E_b/N_0$ operating point, all compared decoders are evaluated using identical payloads and channel realizations. The encoding and decoding pipelines are implemented on top of the Sionna library~\cite{sionna}. {Polar-code construction follows the 5G NR reliability sequence specified in 3GPP TS 38.212 \cite{3gpp}, and CRC-aided decoding uses the CRC-16 polynomial $g(x)=x^{16}+x^{12}+x^5+1$. For each decoder and code rate, $5{,}000$, $5{,}000$, $20{,}000$, and $50{,}000$ blocks were evaluated at SNRs of $0$, $0.5$, $1$, and $1.5$ dB, respectively, while $100{,}000$ blocks were evaluated at each SNR of $2$ dB or higher.}

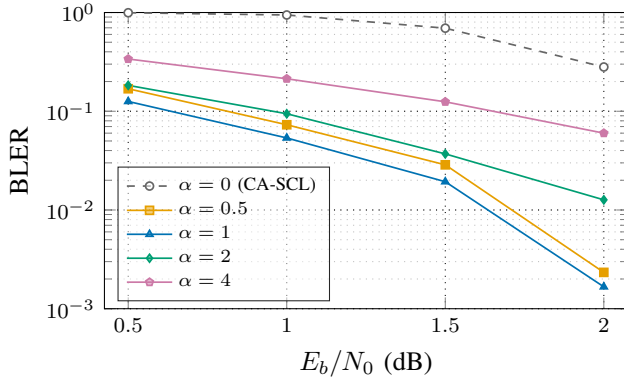
\begin{figure}[t]
\centering
\begin{tikzpicture}
\definecolor{curve1}{RGB}{0,114,178}
\definecolor{curve2}{RGB}{0,158,115}
\definecolor{curve3}{RGB}{102,102,102}
\definecolor{curve4}{RGB}{230,159,0}
\definecolor{curve5}{RGB}{204,121,167}
\begin{axis}[
    width=0.96\linewidth, height=5.5cm,
    xlabel={$E_b/N_0$ (dB)}, ylabel={BLER},
    ymode=log, log basis y=10, ymin=0.001, ymax=1,
    xmin=0.425, xmax=2.075, xtick={0.5,1,1.5,2},
    tick label style={font=\footnotesize},
    xmajorgrids, ymajorgrids, yminorgrids,
    major grid style={dotted,black}, minor grid style={dotted},
    legend cell align=left,
    legend style={at={(0.025,0.025)}, anchor=south west, align=left,
    draw=white!15!black, font=\scriptsize, row sep=-1pt, legend columns=1,
    fill opacity=0.5, text opacity=1},
    unbounded coords=jump
]
\addplot[line width=0.64pt, curve3, dashed, mark=*, mark size=1.5, mark options={solid,fill=white}] coordinates {
    (0.5, 0.994666666666667) (1, 0.944) (1.5, 0.694) (2, 0.280666666666667)
};
\addlegendentry{$\alpha=0$ (CA-SCL)}
\addplot[line width=0.64pt, curve4, mark=square*, mark size=1.5, mark options={solid}] coordinates {
    (0.5, 0.169) (1, 0.073) (1.5, 0.0286666666666667) (2, 0.00233333333333333)
};
\addlegendentry{$\alpha=0.5$}
\addplot[line width=0.64pt, curve1, mark=triangle*, mark size=1.5, mark options={solid}] coordinates {
    (0.5, 0.125666666666667) (1, 0.0536666666666667) (1.5, 0.0193333333333333) (2, 0.00166666666666667)
};
\addlegendentry{$\alpha=1$}
\addplot[line width=0.64pt, curve2, mark=diamond*, mark size=1.5, mark options={solid}] coordinates {
    (0.5, 0.183) (1, 0.0943333333333333) (1.5, 0.037) (2, 0.0126666666666667)
};
\addlegendentry{$\alpha=2$}
\addplot[line width=0.64pt, curve5, mark=pentagon*, mark size=1.5, mark options={solid}] coordinates {
    (0.5, 0.338333333333333) (1, 0.213333333333333) (1.5, 0.124666666666667) (2, 0.06)
};
\addlegendentry{$\alpha=4$}
\end{axis}
\end{tikzpicture}
\vspace{-0.5em}
\caption{BLER of Semantic SCL for different $\alpha$, with $N=512$, $R=0.75$, and $L=8$. The case $\alpha=0$ is conventional CA-SCL.}
\label{fig:alpha_calibration}
\end{figure}

We select the weighting parameter $\alpha$ by evaluating the $N=512$, $R=0.75$ polar code with $L=8$ and $\alpha\in\{0,0.5,1,2,4\}$ across $E_b/N_0$ from $0.5$ to $2$~dB {on the validation set.} As shown in Fig.~\ref{fig:alpha_calibration}, $\alpha=1$ achieves lower BLER than both smaller and larger values of $\alpha$ at every $E_b/N_0$. This agrees with Section~\ref{sec:lm:bitcost}, where $\alpha=1$ yields the maximum a posteriori rule. We therefore fix $\alpha=1$ for all subsequent test evaluations, although this value is not separately optimized for different code rates or block lengths.

\vspace{-0.3em}
\subsection{BLER Analysis}\label{sec:exp:bler}
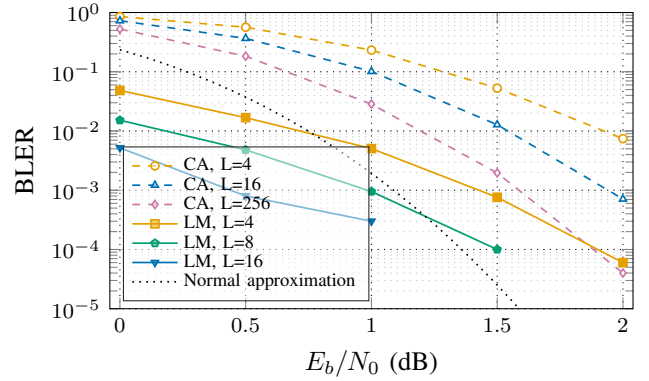
\begin{figure}[t]
\centering
\begin{tikzpicture}
\definecolor{curve1}{RGB}{0,114,178}
\definecolor{curve2}{RGB}{0,158,115}
\definecolor{curve3}{RGB}{230,159,0}
\definecolor{curve4}{RGB}{204,121,167}
\begin{axis}[
    width=0.96\linewidth, height=5.5cm,
    xlabel={$E_b/N_0$ (dB)}, ylabel={BLER},
    ymode=log, log basis y=10, ymin=1e-05, ymax=1,
    xmin=-0.04, xmax=2.04, xtick={0,0.5,1,1.5,2},
    tick label style={font=\footnotesize},
    xmajorgrids, ymajorgrids, yminorgrids,
    major grid style={dotted,black}, minor grid style={dotted},
    legend cell align=left,
    legend style={at={(0.025,0.025)}, anchor=south west, align=left,
    draw=white!15!black, font=\scriptsize, row sep=-3pt, legend columns=1,
    fill opacity=0.5, text opacity=1},
    unbounded coords=jump
]
\addplot[line width=0.64pt, curve3, dashed, mark=*, mark size=1.5, mark options={solid,fill=white}] coordinates {
    (0, 0.8544) (0.5, 0.5644) (1, 0.23215) (1.5, 0.0529) (2, 0.00743)
};
\addlegendentry{CA, L=4}
\addplot[line width=0.64pt, curve1, dashed, mark=triangle*, mark size=1.5, mark options={solid,fill=white}] coordinates {
    (0, 0.7266) (0.5, 0.3658) (1, 0.1012) (1.5, 0.01276) (2, 0.00071)
};
\addlegendentry{CA, L=16}
\addplot[line width=0.64pt, curve4, dashed, mark=diamond*, mark size=1.5, mark options={solid,fill=white}] coordinates {
    (0, 0.5258) (0.5, 0.183) (1, 0.0284) (1.5, 0.00196) (2, 4e-05)
};
\addlegendentry{CA, L=256}
\addplot[line width=0.64pt, curve3, mark=square*, mark size=1.5, mark options={solid}] coordinates {
    (0, 0.0484) (0.5, 0.0168) (1, 0.00505) (1.5, 0.00076) (2, 6e-05)
};
\addlegendentry{LM, L=4}
\addplot[line width=0.64pt, curve2, mark=pentagon*, mark size=1.5, mark options={solid}] coordinates {
    (0, 0.0152) (0.5, 0.0048) (1, 0.00095) (1.5, 0.0001) (2, nan)
};
\addlegendentry{LM, L=8}
\addplot[line width=0.64pt, curve1, mark=triangle*, mark size=1.5, mark options={solid,rotate=180}] coordinates {
    (0, 0.0052) (0.5, 0.0008) (1, 0.0003) (1.5, nan) (2, nan)
};
\addlegendentry{LM, L=16}
\addplot[line width=0.64pt, black, dotted] coordinates {(0, 0.234572578577206) (0.05, 0.20383823481365) (0.1, 0.175543083719848) (0.15, 0.149782772963262) (0.2, 0.126593611761749) (0.25, 0.105956179186619) (0.3, 0.087800908766116) (0.35, 0.0720152102223053) (0.4, 0.0584516306629409) (0.45, 0.0469365387367548) (0.5, 0.0372788345604216) (0.55, 0.0292782407344602) (0.6, 0.0227328080073474) (0.65, 0.0174453640455327) (0.7, 0.0132287357095432) (0.75, 0.00990967509658085) (0.8, 0.00733150960805026) (0.85, 0.00535561061536323) (0.9, 0.0038618303869466) (0.95, 0.00274809156437033) (1, 0.00192932846579294) (1.05, 0.00133597732219567) (1.1, 0.000912196790990954) (1.15, 0.000613974853308736) (1.2, 0.000407247610541304) (1.25, 0.00026612327014239) (1.3, 0.00017127377091955) (1.35, 0.000108529218617778) (1.4, 6.76878908223472e-05) (1.45, 4.15375474652703e-05) (1.5, 2.50720341774506e-05) (1.55, 1.48801208989451e-05) (1.6, 8.68033875376468e-06) (1.65, 4.97532002373484e-06) (1.7, 2.80088811543365e-06) (1.75, 1.54807646468268e-06) (1.8, 8.39725853410791e-07) (1.85, 4.46840477694961e-07) (1.9, 2.33159638604108e-07) (1.95, 1.19247843297362e-07) (2, 5.97514652569851e-08)};
\addlegendentry{Normal approximation}
\end{axis}
\end{tikzpicture}
\vspace{-0.5em}
\caption{BLER for $N=512$ and $R=0.5$. CA and LM denote conventional and Semantic SCL. 
}
\label{fig:n512_r050_bler}
\end{figure}

\begin{figure}[t]
\centering
\begin{tikzpicture}
\definecolor{curve1}{RGB}{0,114,178}
\definecolor{curve2}{RGB}{0,158,115}
\definecolor{curve3}{RGB}{230,159,0}
\definecolor{curve4}{RGB}{204,121,167}
\begin{axis}[
    width=0.96\linewidth, height=5.5cm,
    xlabel={$E_b/N_0$ (dB)}, ylabel={BLER},
    ymode=log, log basis y=10, ymin=1e-05, ymax=1,
    xmin=-0.04, xmax=4.04, xtick={0,0.5,1,1.5,2,2.5,3,3.5,4},
    tick label style={font=\footnotesize},
    xmajorgrids, ymajorgrids, yminorgrids,
    major grid style={dotted,black}, minor grid style={dotted},
    legend cell align=left,
    legend style={at={(0.025,0.025)}, anchor=south west, align=left,
    draw=white!15!black, font=\scriptsize, row sep=-1pt, legend columns=1,
    fill opacity=0.5, text opacity=1},
    unbounded coords=jump
]
\addplot[line width=0.64pt, curve3, dashed, mark=*, mark size=1.5, mark options={solid,fill=white}] coordinates {
    (0, 1) (0.5, 0.9982) (1, 0.962) (1.5, 0.77782) (2, 0.39496) (2.5, 0.10255) (3, 0.01223) (3.5, 0.00078) (4, 3e-05)
};
\addlegendentry{CA, L=4}
\addplot[line width=0.64pt, curve1, dashed, mark=triangle*, mark size=1.5, mark options={solid,fill=white}] coordinates {
    (0, 0.9998) (0.5, 0.9936) (1, 0.91725) (1.5, 0.63102) (2, 0.22683) (2.5, 0.03374) (3, 0.00182) (3.5, 2e-05) (4, nan)
};
\addlegendentry{CA, L=16}
\addplot[line width=0.64pt, curve4, dashed, mark=diamond*, mark size=1.5, mark options={solid,fill=white}] coordinates {
    (0, 0.9994) (0.5, 0.9824) (1, 0.82485) (1.5, 0.42682) (2, 0.09084) (2.5, 0.00586) (3, 8e-05) (3.5, nan) (4, nan)
};
\addlegendentry{CA, L=256}
\addplot[line width=0.64pt, curve3, mark=square*, mark size=1.5, mark options={solid}] coordinates {
    (0, 0.4084) (0.5, 0.2634) (1, 0.1275) (1.5, 0.04636) (2, 0.01208) (2.5, 0.00216) (3, 0.00021) (3.5, nan) (4, nan)
};
\addlegendentry{LM, L=4}
\addplot[line width=0.64pt, curve2, mark=pentagon*, mark size=1.5, mark options={solid}] coordinates {
    (0, 0.2038) (0.5, 0.1102) (1, 0.0428) (1.5, 0.01248) (2, 0.00275) (2.5, 0.00035) (3, 1e-05) (3.5, nan) (4, nan)
};
\addlegendentry{LM, L=8}
\addplot[line width=0.64pt, curve1, mark=triangle*, mark size=1.5, mark options={solid,rotate=180}] coordinates {
    (0, 0.0962) (0.5, 0.0406) (1, 0.0143) (1.5, 0.00368) (2, 0.00068) (2.5, 7e-05) (3, nan) (3.5, nan) (4, nan)
};
\addlegendentry{LM, L=16}
\addplot[line width=0.64pt, black, dotted] coordinates {(0, 0.993743070338621) (0.05, 0.991523642715747) (0.1, 0.988632455352241) (0.15, 0.984908422354663) (0.2, 0.980165780867861) (0.25, 0.974194574946684) (0.3, 0.966762634337306) (0.35, 0.957619350482695) (0.4, 0.946501491270528) (0.45, 0.933141194167841) (0.5, 0.917276133682504) (0.55, 0.898661677740952) (0.6, 0.87708463855327) (0.65, 0.852378002864434) (0.7, 0.824435815618532) (0.75, 0.793227215320464) (0.8, 0.758808505437187) (0.85, 0.72133211871486) (0.9, 0.681051409198795) (0.95, 0.638320399438045) (1, 0.593587914720243) (1.05, 0.547385934859896) (1.1, 0.500312455579676) (1.15, 0.453009632537173) (1.2, 0.406138430709216) (1.25, 0.360351367856874) (1.3, 0.31626517641885) (1.35, 0.274435279151806) (1.4, 0.235333864037987) (1.45, 0.199333058855292) (1.5, 0.166694272733892) (1.55, 0.137564237422805) (1.6, 0.111977704407668) (1.65, 0.0898662000924897) (1.7, 0.0710717707732342) (1.75, 0.0553643106841542) (1.8, 0.0424608899974137) (1.85, 0.0320454933342159) (1.9, 0.0237877292511284) (1.95, 0.0173593450009544) (2, 0.0124477339146011) (2.05, 0.00876600531402166) (2.1, 0.00605955177892514) (2.15, 0.0041093575120153) (2.2, 0.00273251913396494) (2.25, 0.00178058572853365) (2.3, 0.00113637126225356) (2.35, 0.000709863324882037) (2.4, 0.000433767924372243) (2.45, 0.000259113701183312) (2.5, 0.000151211910213044) (2.55, 8.61482545974056e-05) (2.6, 4.78809894087208e-05) (2.65, 2.59427270717657e-05) (2.7, 1.3692145023851e-05) (2.75, 7.0337222496025e-06) (2.8, 3.51396632154132e-06) (2.85, 1.70582682048976e-06) (2.9, 8.039115296096e-07) (2.95, 3.67461989214799e-07) (3, 1.62751603634641e-07) (3.05, 6.97761035747617e-08) (3.1, 2.89266207541588e-08) (3.15, 1.15829521162019e-08) (3.2, 4.47477695173397e-09) (3.25, 1.66583486131287e-09) (3.3, 5.96836577863631e-10) (3.35, 2.05528361571664e-10) (3.4, 6.79336460221579e-11) (3.45, 2.15214533704714e-11) (3.5, 6.52498708551942e-12) (3.55, 1.89028212622892e-12) (3.6, 5.22393818654169e-13) (3.65, 1.37481932801247e-13) (3.7, 3.43942303709008e-14) (3.75, 8.16387868544889e-15) (3.8, 1.83492134270706e-15) (3.85, 3.89713526304617e-16) (3.9, 7.80429398789987e-17) (3.95, 1.47024001745203e-17) (4, 2.59936005887728e-18)};
\addlegendentry{Normal approximation}
\end{axis}
\end{tikzpicture}
\vspace{-0.5em}
\caption{BLER for $N=512$ and $R=0.75$. CA and LM denote conventional and Semantic SCL.}
\label{fig:n512_r075_bler}
\end{figure}
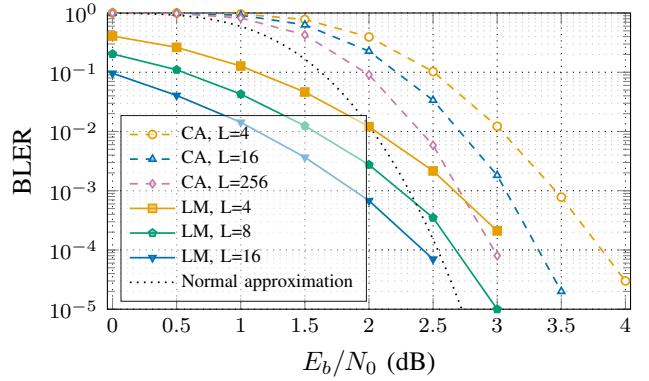

\vspace{-0.3em}
Semantic SCL substantially reduces BLER for both code rates as seen in Fig.~\ref{fig:n512_r050_bler} and Fig.~\ref{fig:n512_r075_bler}. For $R=0.75$ and $L=16$, the BLER decreases from $6.31\times10^{-1}$ to $3.68\times10^{-3}$ at $E_b/N_0=1.5$~dB and from $3.37\times10^{-2}$ to $7.00\times10^{-5}$ at $2.5$~dB, corresponding to reductions of approximately $171\times$ and $482\times$, respectively. At the same rate, Semantic SCL with $L=4$ has lower BLER than conventional CA-SCL with $L=256$ up to $2.5$~dB. For $R=0.5$, $L=16$, and $E_b/N_0=1$~dB, BLER decreases from $1.01\times10^{-1}$ to $3.00\times10^{-4}$. At this rate, Semantic SCL with $L=4$ outperforms CA-SCL with $L=256$ up to $1.5$~dB. These results show that language-model guidance substantially improves decoding at matched list sizes and can achieve lower BLER than conventional decoding with much larger lists. Semantic SCL with $L=16$ lies below the normal approximation for $8B$ uniformly distributed bits~\cite{PPV2010l}, because the payload is not uniformly distributed.

\vspace{-0.3em}
\subsection{Latency}\label{sec:exp:latency}
\vspace{-0.3em}
We report the amortized decoding time per block, namely the synchronized decoding time of a batch divided by the batch size. Timing includes polar decoding, LM inference, and CRC selection, and excludes warm-up, model loading, encoding, noise generation, and file I/O. These timings refer to our software implementation and need not reflect the relative cost of dedicated hardware decoders. {CA-SCL with $L\geq256$ uses batch size 4 to limit the memory required for concurrent decoding paths; all other configurations use batch size 64.} 

Table~\ref{tab:runtime} compares the average decoding latency per block. Incorporating the language-model prior increases latency relative to conventional CA-SCL decoding at the same list size. For $R=0.5$, Semantic SCL requires $5.497$~ms and $14.494$~ms for $L=4$ and $L=16$, respectively, compared with $1.944$~ms and $3.471$~ms for conventional CA-SCL. A similar trend is observed for $R=0.75$, where the corresponding runtimes increase from $2.107$~ms to $9.559$~ms for $L=4$ and from $3.851$~ms to $29.682$~ms for $L=16$. The overhead is dominated by the $B$ batched forward passes of the language model, each over at most $L$ paths, whereas the prefix masses in~\eqref{eq:mass} add only $O(BL)$ operations.

Semantic SCL with $L=8$ outperforms conventional CA-SCL with $L=256$ in Section~\ref{sec:exp:bler}, while requiring considerably lower latency ($8.178$ versus $55.689$~ms for $R=0.5$ and $16.291$ versus $61.951$~ms for $R=0.75$).  Thus, the additional language-model computation can partly replace the cost of maintaining substantially larger SCL lists. A threshold-based approach can further reduce the number of LM evaluations and hence the latency, at the cost of guiding fewer bytes, as examined in Section~\ref{sec:exp:threshold}.

\begin{table}[t]
\vspace{0.05in}
\centering
\caption{Amortized decoding time per block, averaged over $E_b/N_0=0,0.5,\ldots,2$~dB.}
\label{tab:runtime}
\begin{tabular}{lcc}
\hline
\multirow{2}{*}{\textbf{Decoder}} & \textbf{$R=0.5$} & \textbf{$R=0.75$} \\
& \multicolumn{2}{c}{\textbf{Average runtime (ms/block)}} \\
\hline
CA-SCL $L=4$    & 1.944  & 2.107  \\
CA-SCL $L=16$   & 3.471  & 3.851  \\
CA-SCL $L=256$  & 55.689 & 61.951 \\
Semantic $L=4$  & 5.497  & 9.559  \\
Semantic $L=8$  & 8.178  & 16.291 \\
Semantic $L=16$ & 14.494 & 29.682 \\
\hline
\end{tabular}
\end{table}

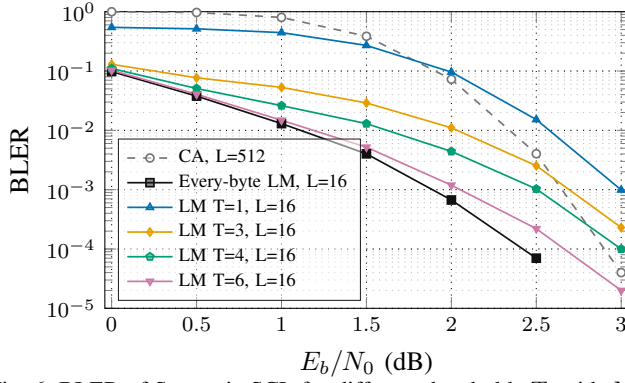
\begin{figure}[t]
\centering
\begin{tikzpicture}
\definecolor{curve1}{RGB}{119,119,119}
\definecolor{curve2}{RGB}{17,17,17}
\definecolor{curve3}{RGB}{0,114,178}
\definecolor{curve4}{RGB}{230,159,0}
\definecolor{curve5}{RGB}{0,158,115}
\definecolor{curve6}{RGB}{204,121,167}
\begin{axis}[
width=0.96\linewidth, height=5.5cm,
xlabel={$E_b/N_0$ (dB)}, ylabel={BLER},
ymode=log, log basis y=10, ymin=1e-05, ymax=1,
xmin=-0.04, xmax=3.04, xtick={0,0.5,1,1.5,2,2.5,3},
tick label style={font=\footnotesize},
xmajorgrids, ymajorgrids, yminorgrids,
major grid style={dotted,black}, minor grid style={dotted},
legend cell align=left,
legend style={at={(0.025,0.025)}, anchor=south west, align=left,
draw=white!15!black, font=\scriptsize, row sep=-1pt, legend columns=1,
fill opacity=0.5, text opacity=1},
unbounded coords=jump
]
\addplot[line width=0.64pt, curve1, dashed, mark=*, mark size=1.5, mark options={solid,fill=white}] coordinates {(0, 0.9982) (0.5, 0.9706) (1, 0.8014) (1.5, 0.38542) (2, 0.07275) (2.5, 0.00403) (3, 4e-05)};
\addlegendentry{CA, L=512}
\addplot[line width=0.64pt, curve2, mark=square*, mark size=1.5, mark options={solid}] coordinates {(0, 0.098) (0.5, 0.0378) (1, 0.013) (1.5, 0.00402) (2, 0.00067) (2.5, 7e-05) (3, nan)};
\addlegendentry{Every-byte LM, L=16}
\addplot[line width=0.64pt, curve3, mark=triangle*, mark size=1.5, mark options={solid}] coordinates {(0, 0.5446) (0.5, 0.5146) (1, 0.4441) (1.5, 0.27014) (2, 0.09603) (2.5, 0.0152) (3, 0.00099)};
\addlegendentry{LM T=1, L=16}
\addplot[line width=0.64pt, curve4, mark=diamond*, mark size=1.5, mark options={solid}] coordinates {(0, 0.13) (0.5, 0.077) (1, 0.05305) (1.5, 0.02898) (2, 0.01106) (2.5, 0.00253) (3, 0.00023)};
\addlegendentry{LM T=3, L=16}
\addplot[line width=0.64pt, curve5, mark=pentagon*, mark size=1.5, mark options={solid}] coordinates {(0, 0.1102) (0.5, 0.0508) (1, 0.0261) (1.5, 0.01298) (2, 0.00442) (2.5, 0.00103) (3, 0.0001)};
\addlegendentry{LM T=4, L=16}
\addplot[line width=0.64pt, curve6, mark=triangle*, mark size=1.5, mark options={solid,rotate=180}] coordinates {(0, 0.1028) (0.5, 0.0406) (1, 0.0148) (1.5, 0.0052) (2, 0.00119) (2.5, 0.00022) (3, 2e-05)};
\addlegendentry{LM T=6, L=16}
\end{axis}
\end{tikzpicture}
\vspace{-0.5em}
\caption{BLER of Semantic SCL for different thresholds $T$, with $N=512$, $R=0.75$ and $L=16$. CA and LM denote conventional and Semantic SCL.}
\label{fig:threshold_bler}
\end{figure}

\vspace{-0.3em}
\subsection{Threshold-based Semantic SCL}\label{sec:exp:threshold}
\vspace{-0.3em}
Fig.~\ref{fig:threshold_bler} and Table~\ref{tab:threshold_latency} illustrate the trade-off between BLER and latency for byte-replay thresholds $T\in\{1,3,4,6\}$, compared with the original every-byte Semantic SCL decoder using $L=16$ and conventional CA-SCL using $L=512$, for $N=512$ and $R=0.75$. Increasing $T$ selects more bytes for LM-guided replay, improving BLER at additional computational cost, while higher noise also increases LM invocation through weaker channel evidence. With $T=6$, the decoding time at $2$~dB is about two thirds lower than with every-byte guidance, while the $E_b/N_0$ required for a BLER of $10^{-3}$ increases by about $0.2$~dB. At $E_b/N_0=2$~dB, reducing $T$ from 6 to 1 decreases decoding time from $9.382$ to $6.180$~ms/block, a reduction of approximately $34\%$. However, sparser guidance increases BLER from $1.19\times10^{-3}$ to $9.60\times10^{-2}$. All evaluated thresholds decode faster than conventional CA-SCL with $L=512$. Thus, under the fixed weight $\alpha=1$, selective guidance reduces computational cost at the expense of error-correction performance, with every-byte guidance achieving the lowest BLER across the evaluated SNR range.

\begin{table}[t]
\vspace{0.05in}
\centering
\caption{Average decoding time per block (ms/block) for threshold-based byte replay, every-byte LM guidance, and conventional CA-SCL.}
\label{tab:threshold_latency}
\footnotesize
\setlength{\tabcolsep}{3pt}
\begin{tabular}{c|cccccc}
    \hline
    $E_b/N_0$ (dB) & $T=1$ & $T=3$ & $T=4$ & $T=6$ & Every-byte & CA-SCL \\
    \hline
    0   & 10.912 & 14.893 & 16.164 & 18.145 & 29.702 & 122.541 \\
    1   & 7.651  & 10.617 & 11.756 & 13.898 & 29.672 & 122.514 \\
    2   & 6.180  & 7.304  & 7.919  & 9.382  & 29.698 & 122.561 \\
    3   & 5.170  & 5.947  & 6.075  & 6.479  & 30.850 & 126.802 \\
    \hline
\end{tabular}

\end{table}

\vspace{-0.3em}
\vspace{-0.3em}
\vspace{-0.2em}
\section{Conclusion}\label{sec:conclusion}
\vspace{-0.3em}
We proposed a Semantic SCL decoder for polar codes that carry natural-language text. The decoder splits the next-byte probability of a byte-level language model into conditional bit costs, which are added to the path metric bit by bit. During the decoding, the path list is pruned based on the resulting metric. For $N=512$ and $R=0.75$, the proposed decoder reduces the BLER of CA-SCL with $L=16$ by a factor of about 171 at $E_b/N_0=1.5$~dB. With $L=8$, it also outperforms CA-SCL with $L=256$ at about one quarter of its decoding time in our software implementation. Invoking the language model only for bytes with unreliable channel LLRs reduces the decoding time by about two thirds, at a loss of about $0.2$~dB.

\bibliographystyle{IEEEtran}
\bibliography{ref/IEEEabrv,ref/OSDAbrv,ref/paper_ref_correct,ref/reference}

\end{document}